\documentclass[onecolumn,secnumarabic,amssymb, nobibnotes, aps, prd]{revtex4-2}

\usepackage{comment}
\usepackage{graphicx}
\usepackage{amsmath}
\usepackage{mathtools}
\usepackage{slashed}
\usepackage{xcolor}
\usepackage[force]{feynmp-auto}
\usepackage{grffile}
\usepackage{float}
\usepackage{bbm}
\usepackage{hyperref}
\usepackage{MnSymbol}

\begin{document}

\title{\Large Axion contribution to the mass-radius relation of neutron stars}%
\author{Momchil Naydenov$^a$\\ Alberto Salvio\,$^{b,c}$ \\
\vspace{0.2cm}
{\it $^a$ Faculty of Physics, Sofia University, 5 J. Bourchier Blvd, 1164 Sofia, Bulgaria \\ 
\vspace{0.2cm}
$^b$ Physics Department, University of Rome Tor Vergata, \\ 
via della Ricerca Scientifica, I-00133 Rome, Italy\\ 
\vspace{0.2cm}$^c$ I.N.F.N. -  Rome Tor Vergata,\\
via della Ricerca Scientifica, I-00133 Rome, Italy}
\vspace{0.4cm}%
}
\maketitle

\section*{Abstract}
\noindent An effective Lagrangian for the interaction between a pseudo-scalar (axion-like) field and massive fermions is considered. At high density and non-zero temperature axions can be produced through bremsstrahlung. If the axion-neutron interaction is greater than 
a certain value we can have a mean free path smaller than the size of a neutron star.
The influence of such trapped axions on the mass-radius function of the neutron star is investigated by solving numerically the Tolman-Oppenheimer-Volkoff equations. We show that causality limits the applicability of the trapping regime. We find specific ranges of central densities and temperatures of a neutron star for which axions give a conspicuous contribution to the neutron-star mass. For other densities and temperatures axions do not have a significant effect on the structure of a neutron star. Since we use an effective approach our results can  be easily applied to specific axion models.

\section{Introduction}
In the Standard Model of elementary particles, it is possible to include in the QCD Lagrangian a CP-violating term
\begin{equation}
\mathcal{L}_\theta = \frac{\theta}{32
\pi^2}\int d^4x\epsilon^{\mu\nu\alpha\beta}~\text{Tr}(G_{\mu\nu}G_{\alpha\beta}),
\end{equation} which remains unobserved to date. Here $G_{\mu\nu}$ is the QCD field strength. To solve  the problem of explaining why strong interactions seem to preserve CP (strong CP problem), as indicated by measurements of the neutron electric dipole moment, in 1977 R. Peccei and H. Quinn published some work~\cite{PQ} suggesting the existence of a chiral Abelian symmetry, U$(1)_\text{PQ}$, which implies that strong interactions do preserve CP.
The spontaneous breaking of U$(1)_\text{PQ}$, which is also broken by anomalies,  results in a Goldstone boson, the QCD axion, as recognized independently by F. Wilczek and S. Weinberg~\cite{will,wein}. In this solution to the strong CP problem, $\theta$ is replaced by the axion field, which we, therefore, denote with the same symbol for simplicity.
 At the same time,
  the QCD axion is a good cold dark matter candidate
~\cite{preskil, sikivie, Fischler}. Light axions and axion-like particles have been included in a wide variety of theoretical frameworks extending the Standard Model in the hope to guide experimental activities that could discover such particles in the future.


    Compact stellar objects seem
    to be a promising stage for testing the limits of the existing theories. The fact that axions can be thermally produced in the Universe~\cite{Masso:2002np,Graf:2010tv,salvio,DEramo:2021psx} and the role of these particles in the cooling of supernovae and neutron stars at finite density~\cite{balkin,springman} suggest that, if they exist, they should be included
    in the model building of compact objects. Coupling 
     axions to photons or fermions may result in various mechanisms for energy loss of newly formed neutron stars~\cite{reffelt, gomez, sedrakian, garbreht}, it may lead to formation of axion clouds around neutron stars~\cite{cloud} and with the advances of observational astrophysics we might be able to access new territories in the parameter space of exotic particles.

In this work we consider a simple model of a Fermi gas at low temperature, comprised of neutrons which interact with thermally generated axions, in the regime where they are trapped inside a neutron star. 
From statistical considerations we derive the equation of state for this ensemble and use it in the Tolman-Oppenheimer-Volkoff (TOV) equations to explore the impact of axions on a neutron stellar gas. The physics of the strong interactions between particles in the interior of a compact object, as well as the presence of different exotic matter states like quark-gluon plasma, is not taken into account. The axion production is influenced by the matter density and temperature, which we take from current observations. The effective Lagrangian includes the free parameters $g$ -- the strength of interaction between the neutron and the axion --  and $M$, the axion mass.
In our simple model 
we show that the axion field can have a significant influence on the mass-radius dependence.

 In this work natural units are adopted, where $\hbar=k_B=c=1.$

\section{The model}
Consider an interaction between a pseudo-scalar field $\theta$ and a fermion  $\psi$. At the leading order in $\theta$ the effective fermion Lagrangian reads~\cite{Georgi:1986df}
\begin{equation}
\overline{\psi}(i\slashed{\partial}-m)\psi+\frac{c_\psi}{F_a}\overline{\psi}\gamma^5\gamma^\mu\psi\partial_\mu\theta,
\label{Lagrangian0}
\end{equation}
where $F_a$ is the axion decay constant. The coupling $c_\psi$ will be treated as a free parameter, but it is computable in the vacuum once the axion model is specified~\cite{GrillidiCortona:2015jxo}. This work is not restricted to a specific axion model and will treat the system in a general manner, where the $\theta$ field is a weakly-interacting light axion-like (ALP) field. Integrating by parts the second term in~(\ref{Lagrangian0}) and using the field equations of $\psi$ at the order in $\theta$ we are working, one obtains the equivalent Lagrangian 
\begin{equation}
\mathcal{L}=\overline{\psi}(i\slashed{\partial}-m)\psi+ig\overline{\psi}\gamma^5\psi\theta,
\end{equation}
\noindent where $g=2mc_\psi/F_a$ is the strength of interaction. At large densities, typical of neutron stars, the neutron-axion coupling can be significantly enhanced~\cite{balkin}; we will include this  effect in $c_\psi$. The classic window of $F_a$ in the QCD axion case is 
 $10^8~\text{GeV}\lesssim F_a \lesssim 10^{12}~\text{GeV}$, which should not be taken as sharp bounds, although it certainly gives a plausible range, see e.g.~\cite{Salvio:2015cja}. Furthermore, $c_\psi$ can be rather large at finite density as observed above,  opening the possibility that $g$ can be sizable compatibly with the bounds on $F_a$.
 
 The full effective Lagrangian for this system after radiative corrections are taken into account must include a quadratic and a quartic term
\begin{equation}
\mathcal{L}'\rightarrow\mathcal{L}+\frac{1}{2}(\partial_\mu\theta)(\partial^\mu\theta)-\frac{1}{2}M^2\theta^2-\frac{1}{4}\lambda\theta^4+...\, .
\label{EffL}
\end{equation}
\noindent 
The renormalized parameters -- the ALP mass $M$ and the interaction parameter $\lambda$ in (\ref{EffL}) -- even if not initially present at the classical level they are anyway dynamically generated from the interactions of the ALP field with the fermion through loop diagrams with two and four external pseudo-scalar legs. The dots represent other possible $\theta$ self interactions. All $\theta$ interactions are small because $\theta$ is a pseudo-Goldstone boson.

We want to explore the interaction between ALPs and neutrons inside a neutron star. The variation of the $\theta$ self interactions due to the different energy densities in the interior of the star does not introduce a sizable effect to the structure of the star.
The perturbation series expansion neglecting $\mathcal{O}(g)$ terms results in the decoupling between the ALP and the fermion field. We want to explore the thermodynamics of a static spherical self-gravitating gas of radius $R$, comprised of neutrons and thermally produced ALPs, including the effect of $g$. The temperature of interest is~\cite{temperature} $10^2$\,eV $\lesssim  T\lesssim 10^{8}$\,eV and the mass of the ALPs is of the order of or smaller than 1 eV, therefore we work in the regime $M\ll T\ll m$. Under those circumstances, when deriving the free energy, we need to take the low-temperature approximation for neutrons and high-temperature approximation for the ALPs. 

At finite temperature the free energy densities are~\cite{laine} (see also~\cite{Salvio:2024upo}): 
\begin{equation}
f_\theta(T)\approx-\frac{\pi^2T^4}{90}+\frac{M^2T^2}{24}-\frac{M^3T}{12\pi}+\frac{3}{4}\lambda\left(\frac{T^4}{144}-\frac{MT^3}{24\pi}\right)+\mathcal{O}(\lambda^2)+...\, , \label{FEaxion}
\end{equation}
\noindent for the ALPs (which are also assumed in thermal equilibrium, we will comment on this point below) and
\begin{equation}
f_\psi(T)=-T^4\left(\frac{m}{2\pi T}\right)^{3/2}e^{-m/T}
\end{equation}
\noindent for the fermions. The exponential suppression of the fermion free energy at low temperature makes the fermion thermal pressure negligible. As far as the neutrons are concern,  the supporting mechanism is the neutron degeneracy pressure given by~\cite{weinberg}
\begin{equation}\label{NeutronP}
P_{\psi}=\frac{8\pi(2\pi)^2}{15m}\left(\frac{3\rho}{8\pi m}\right)^{5/3},
\end{equation}
\noindent where $\rho$ is the neutron mass density and we have assumed that $\rho$ is much less than the critical density for which the Fermi momentum becomes equal to $m$. We are interested in the effect of the presence of the ALP field in the mass-radius relation of neutron stars. Having its free energy density, we compute  its pressure as $P_\theta=-f_\theta.$ 

\begin{figure}[t!] 
    \centering
    \includegraphics[width=0.3\textwidth]{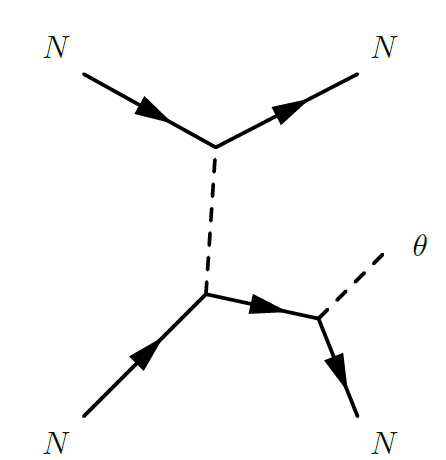} \hspace{2cm} \includegraphics[width=0.4\textwidth]{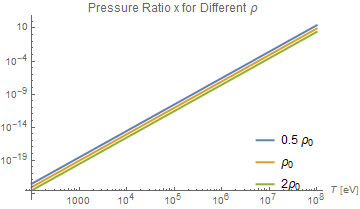}
        \caption{{\bf Left:} Bremsstrahlung emission of an axion-like particle through scattering of neutrons, $N$. {\bf Right:} Ratio $x=P_\theta/P_\psi$ between the axion pressure and the neutron degeneracy pressure over the temperature range $10^2$eV and $10^8$eV. This plot has been done for three different central densities expressed in units of the nuclear density $\rho_0$.}\label{Bremsstrahlung}
 \end{figure}

The main mechanism for ALPs production is the bremsstrahlung process with the corresponding Feynman diagram shown in the left plot of Fig.~\ref{Bremsstrahlung}.
At finite density and  temperature the free-streaming regime is characterized by an energy loss rate due to ALPs emission per unit volume given by~\cite{iwamoto}
\begin{equation}
\epsilon\approx\frac{31}{1890\pi}g^2 m_\pi^{-4} m^{*2}p_F(n)T^6,
\end{equation}
\noindent where $m_\pi$ is the neutral-pion mass, $m^*=0.8m$ and $p_F(n)$ is the Fermi momentum of the neutrons. 
For baryon number density $n_\psi$ the Fermi momentum is fixed by the relation $p_F(n)=340(n_\psi/n_0)^{1/3}~$MeV, where $n_0$ is the nuclear density given by $n_0=1.7\cdot10^{38}~$cm$^{-3}$. The energy of each ALP is $E_\theta\sim T\gg M$, which makes them ultra-relativistic. So, we can estimate the number of axions in the  star  in the free-streaming regime as
\begin{equation}
N_\theta\sim\frac{62}{2835}\pi g^2 m_\pi^{-4} m^{*2}p_F(n)T^5 R^4.
\end{equation}

In this work we discuss a trapping regime~\cite{Burrows:1990pk}, where the mean free path of axions is lower than the neutron-star size. In this situation, axions are mainly produced and absorbed inside the star through bremsstrahlung in the left plot of Fig.~\ref{Bremsstrahlung} and its inverse process, where $\theta$ is in the initial rather than the final state, respectively. This occurs for an axion mass times $c_\psi$ above the $10^{-2}$ eV scale. However, as already pointed out, $c_\psi$ is model dependent and, also, can be significantly enhanced through finite-density effects~\cite{balkin}. So large values of $c_\psi$ (and thus sizable values of $g$) are also possible compatibly with the lower observational bounds on $F_a$. In that case the production and absorption of axions inside the neutron star eventually balance each other  and an  equilibrium number of axions is reached inside the  star. This situation can  be described by an equilibrium (constant in time) density matrix with a thermal distribution of axions~\cite{Burrows:1990pk}.

\section{Mass-radius dependence of a neutron star}

Here we explore the influence of axions on the structure of neutron stars in the trapping regime.
In order to explore the mass-radius dependence of a static neutron star we need to solve numerically the TOV equations:
\begin{equation}
\begin{gathered}
\frac{dP}{dr}=-G\frac{M^*(r)}{r^2}\rho\left(1+\frac{P}{\rho}\right)\left(1+\frac{4\pi r^3P}{M^*(r)}\right)\left(1-\frac{2GM^*(r)}{r}\right)^{-1},\\
\frac{dM^*(r)}{dr}=4\pi r^2\rho.
\end{gathered}
\end{equation}
Here $G$ is the Newton's constant of gravity, $P\approx P_\theta+P_\psi$ is the total pressure of the system 
and $M^*(r)$ is the mass-radius function of the star. For the temperature range of interest $10^2$\,eV $\lesssim  T\lesssim 10^{8}$\,eV, the quantities  $M$ and $\lambda$ have no sizable effect to the axion pressure because $M\ll T$ and $\lambda\ll1$ (see Eq.~(\ref{FEaxion})). The same is true for all $\theta$ self interactions. This leaves us with two free parameters -- the temperature of the neutron star and its central density, defining the initial condition for the TOV equations. We require the central density to be  much less than the critical density such that~(\ref{NeutronP}) holds. In particular, we will illustrate in detail the case of having the central density equal to $\rho(0)=0.5\rho_0$, $\rho(0)=\rho_0$ and $\rho(0)=2\rho_0$, where $\rho_0$ is the nuclear density. The solutions that we explore must be gravitationally stable and causal, so we must have the Jeans length larger than the radius of the star and the speed of sound smaller than the speed of light. The definitions for the Jeans length $\lambda_J$ and the speed of sound $v_s$ are

\begin{equation}
    \lambda_J\equiv v_s\sqrt{\frac{1}{4\pi G \rho}},~v_s\equiv\sqrt{\frac{\partial   P}{\partial\rho}}.
\end{equation}

 Using (\ref{FEaxion}) and~(\ref{NeutronP}) we can plot the ratio between the pressure of the axions over the degeneracy pressure of the neutrons $x=P_\theta/P_\psi$: see the right plot in Fig.~\ref{Bremsstrahlung}.
In that plot one can see that there is a some  temperature range ($T>10^6$~eV) for which the axions could have a sizable effect on the structure of the neutron star. 

This region is  narrowed down from above due to causality: for  high axion pressure, which occurs at high temperature,  the speed of sound may exceed the speed of light. We find that the maximal temperature at which axions would be bound and produce causal effects is $T_{\rm{max}}\approx7.8\cdot10^7$~eV. 
 This temperature is obtained by exploring the speed of sound at different densities and temperatures. The highest causally admissible temperature depends weakly on the central density. 
 
 We also find that the Jeans-length criterion is less restrictive: all causal solutions correspond to gravitationally stable objects. 
 
\begin{figure}[t] 
    \centering
    \includegraphics[width=0.3\textwidth]{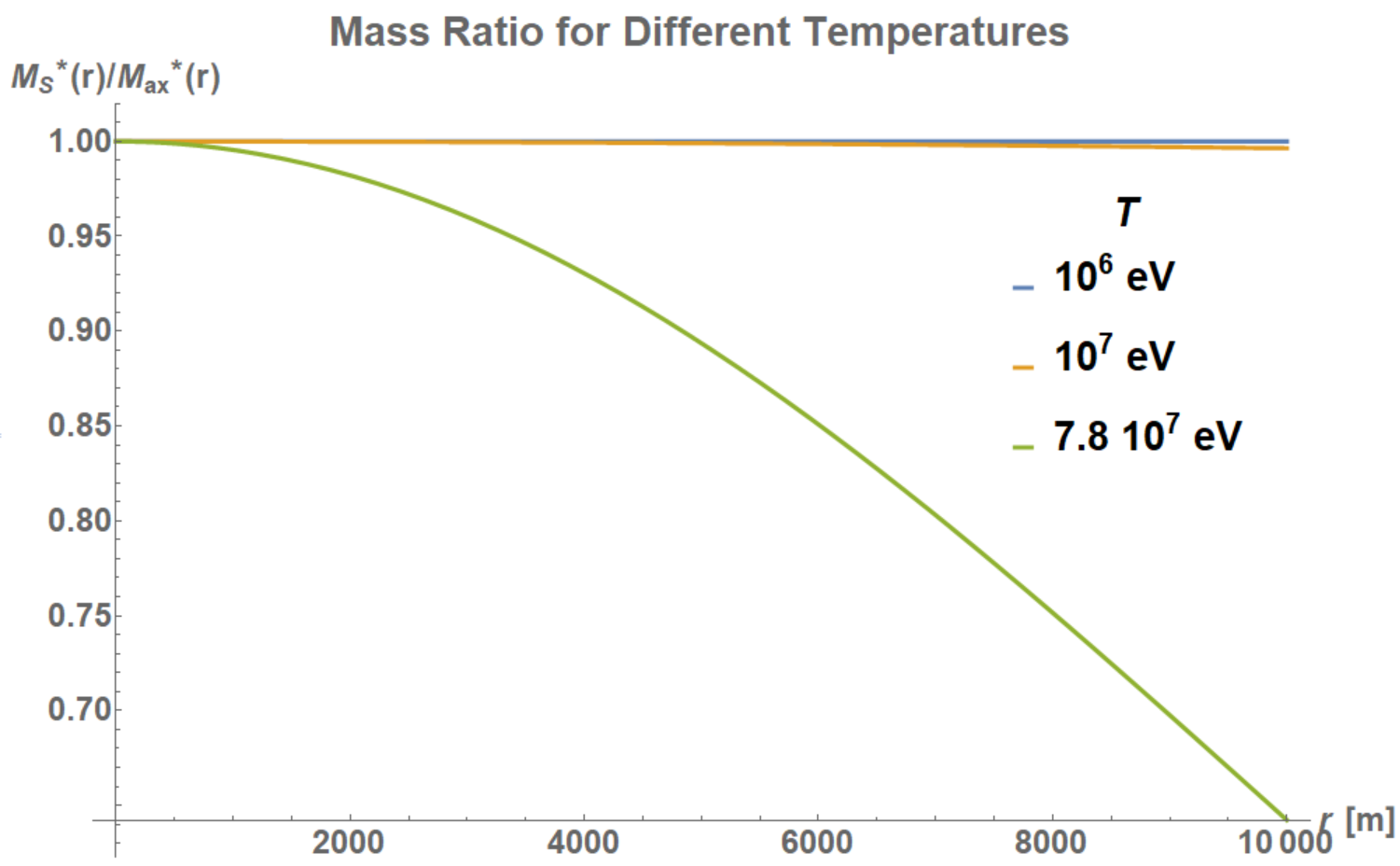}
    \hspace{5mm}
     \includegraphics[width=0.3\textwidth]{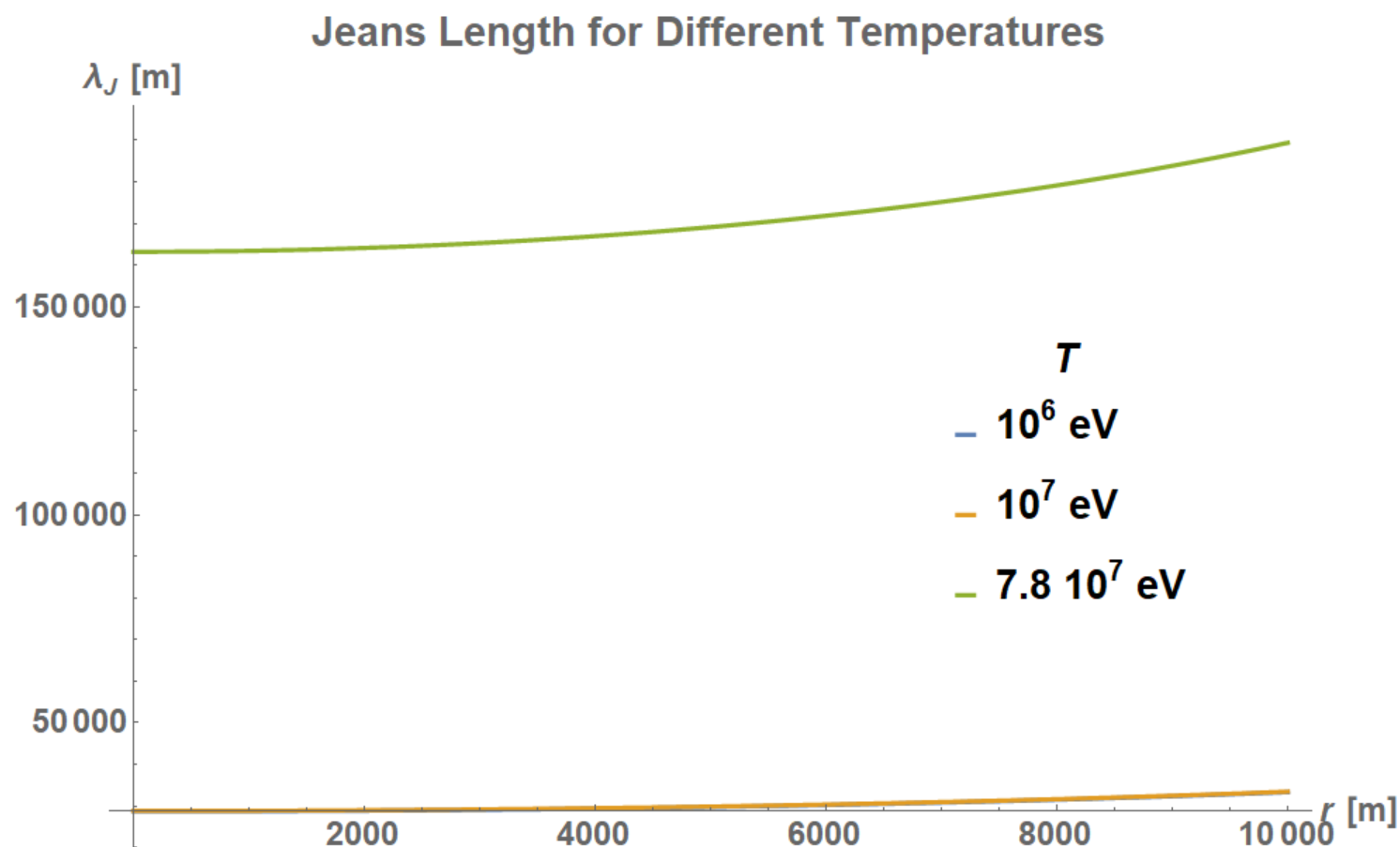}
    \hspace{5mm}
     \includegraphics[width=0.3\textwidth]{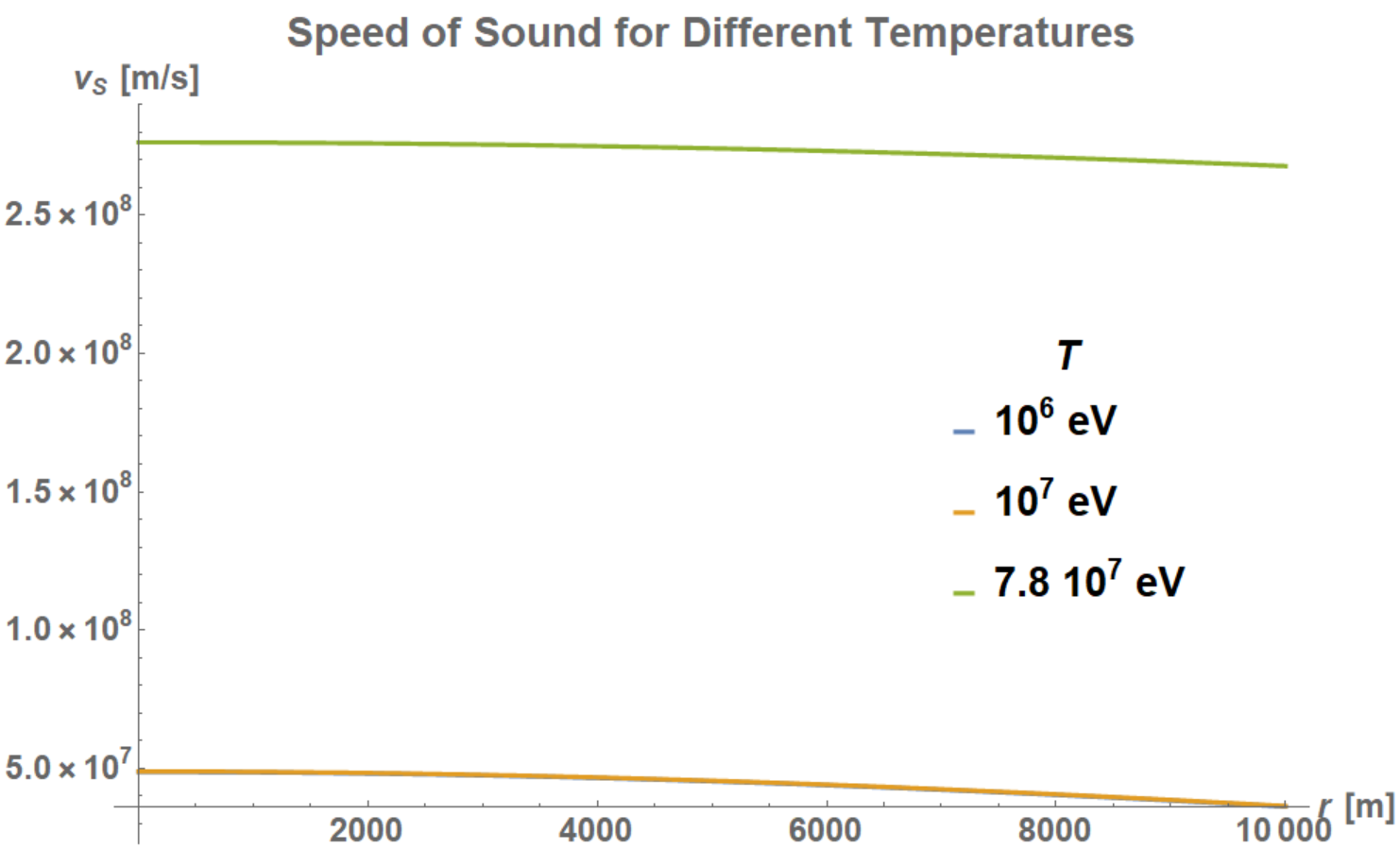}
        \caption{The plots show the ratio of the mass functions of the bare neutron star $M_S^*(r)$ over the same function including axion contribution $M_{\rm{ax}}^*(r)$, the Jeans length and the speed of sound in the medium, all as a function of the radial distance. The  three lines correspond to the temperatures $10^6$~eV, $10^7$~eV and $T_{\rm{max}}$ (blue, orange and green line respectively) while the central density is $0.5\rho_0$ here. It is notable that the axion influence is most sizable at temperatures close to the causal limit. The blue and the orange line are almost coincidental.} \label{05rho0}
 \end{figure}

\begin{figure}[t] 
    \centering
    \includegraphics[width=0.3\textwidth]{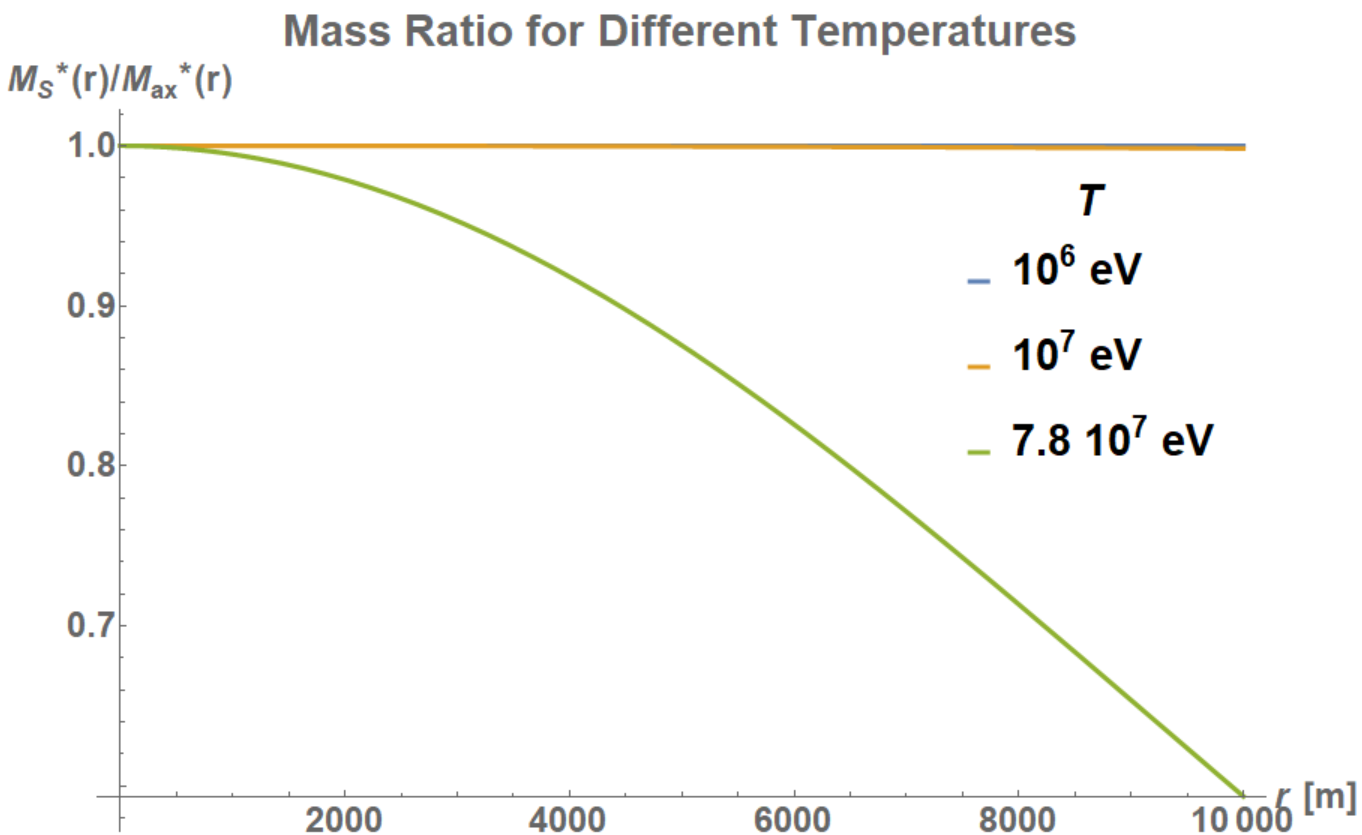}
    \hspace{5mm}
     \includegraphics[width=0.3\textwidth]{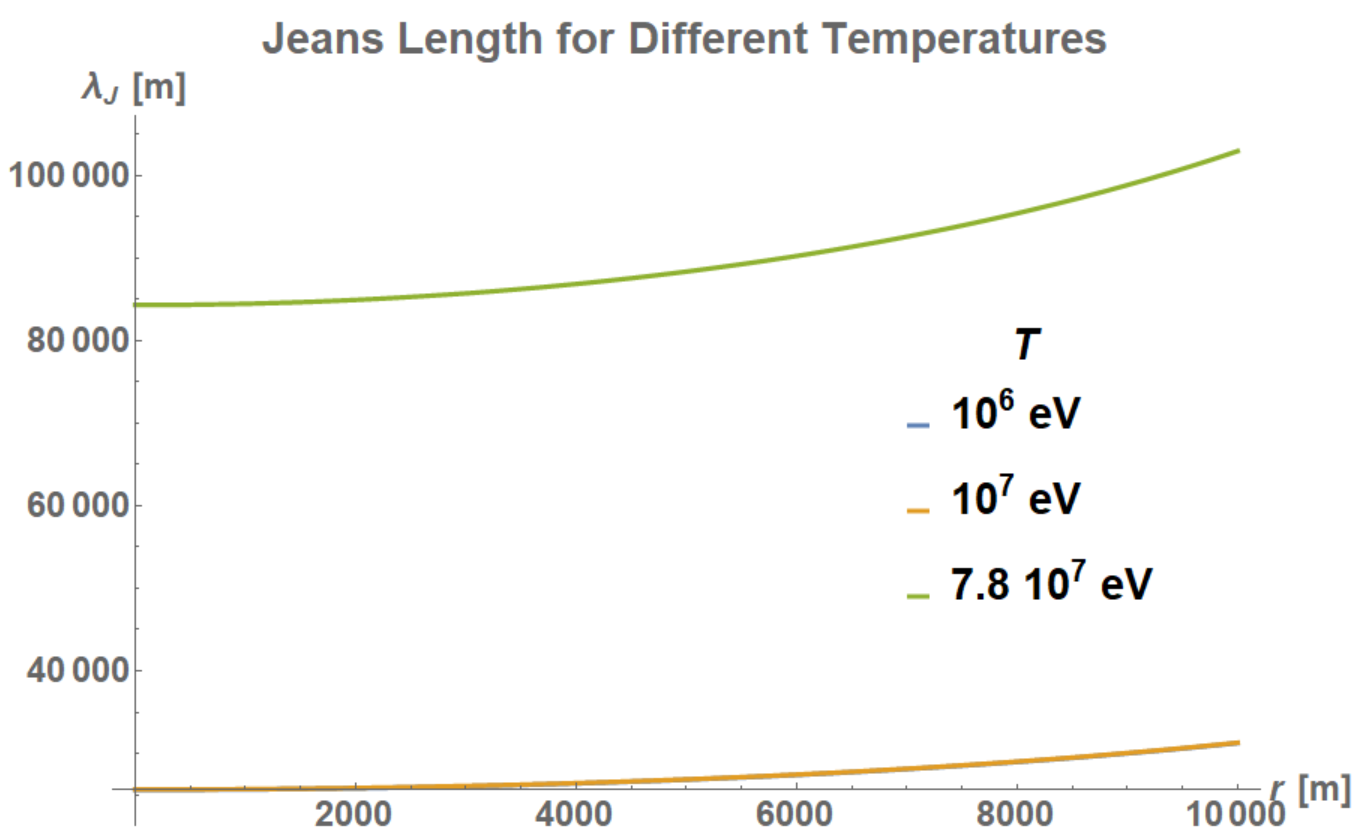}
    \hspace{5mm}
     \includegraphics[width=0.3\textwidth]{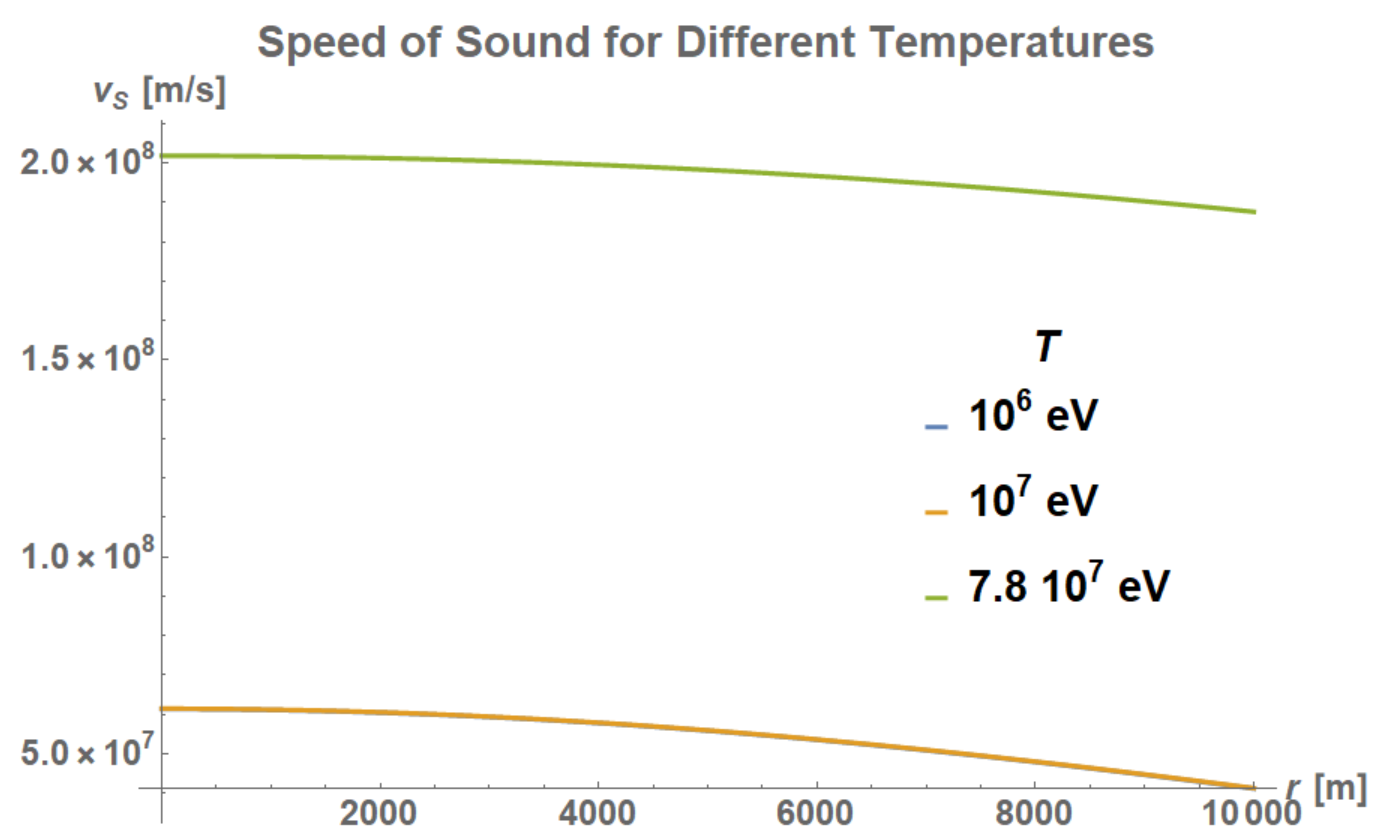}
        \caption{Like in Fig.~\ref{05rho0}, but keeping the central density equal to $\rho_0$. } \label{rho0}
 \end{figure}

\begin{figure}[t] 
    \centering
    \includegraphics[width=0.3\textwidth]{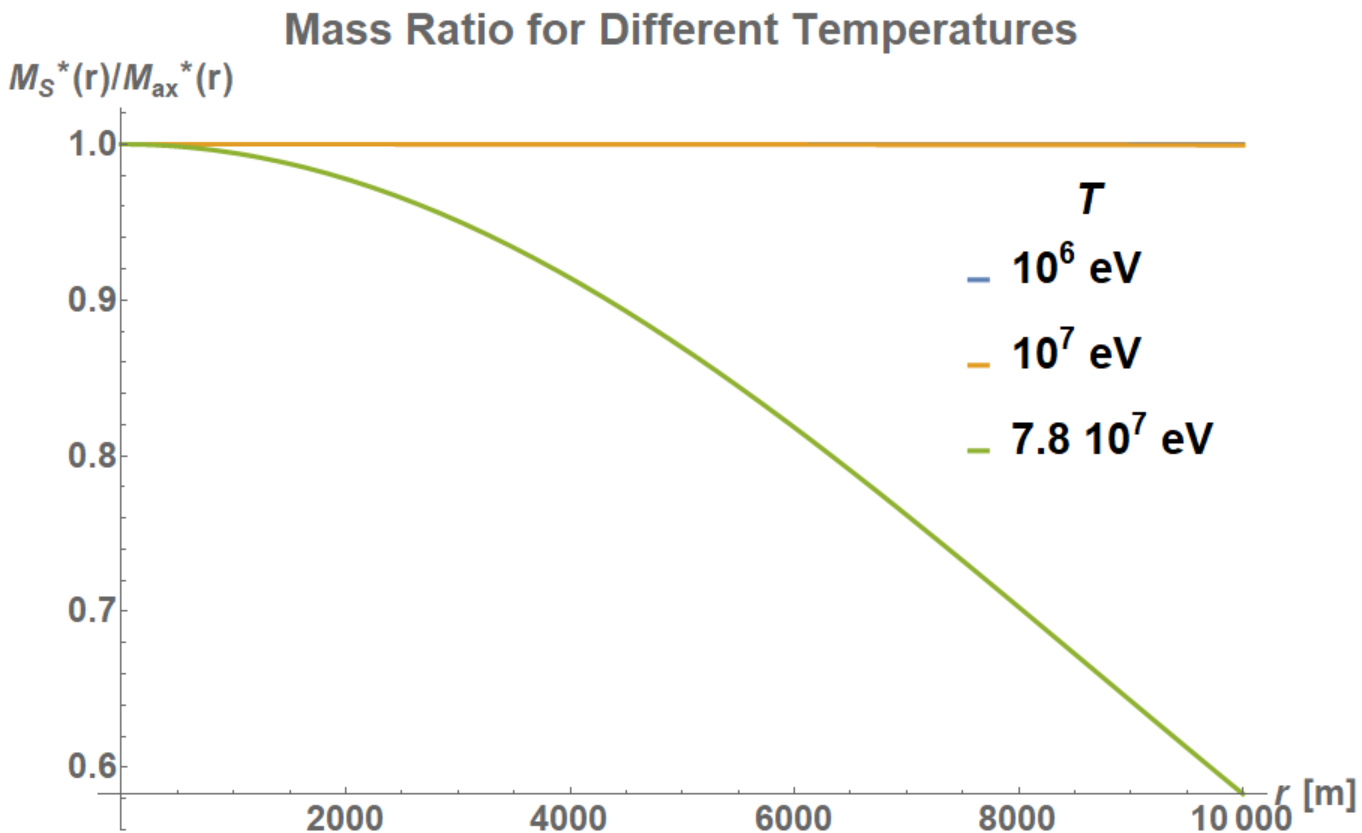}
    \hspace{5mm}
     \includegraphics[width=0.3\textwidth]{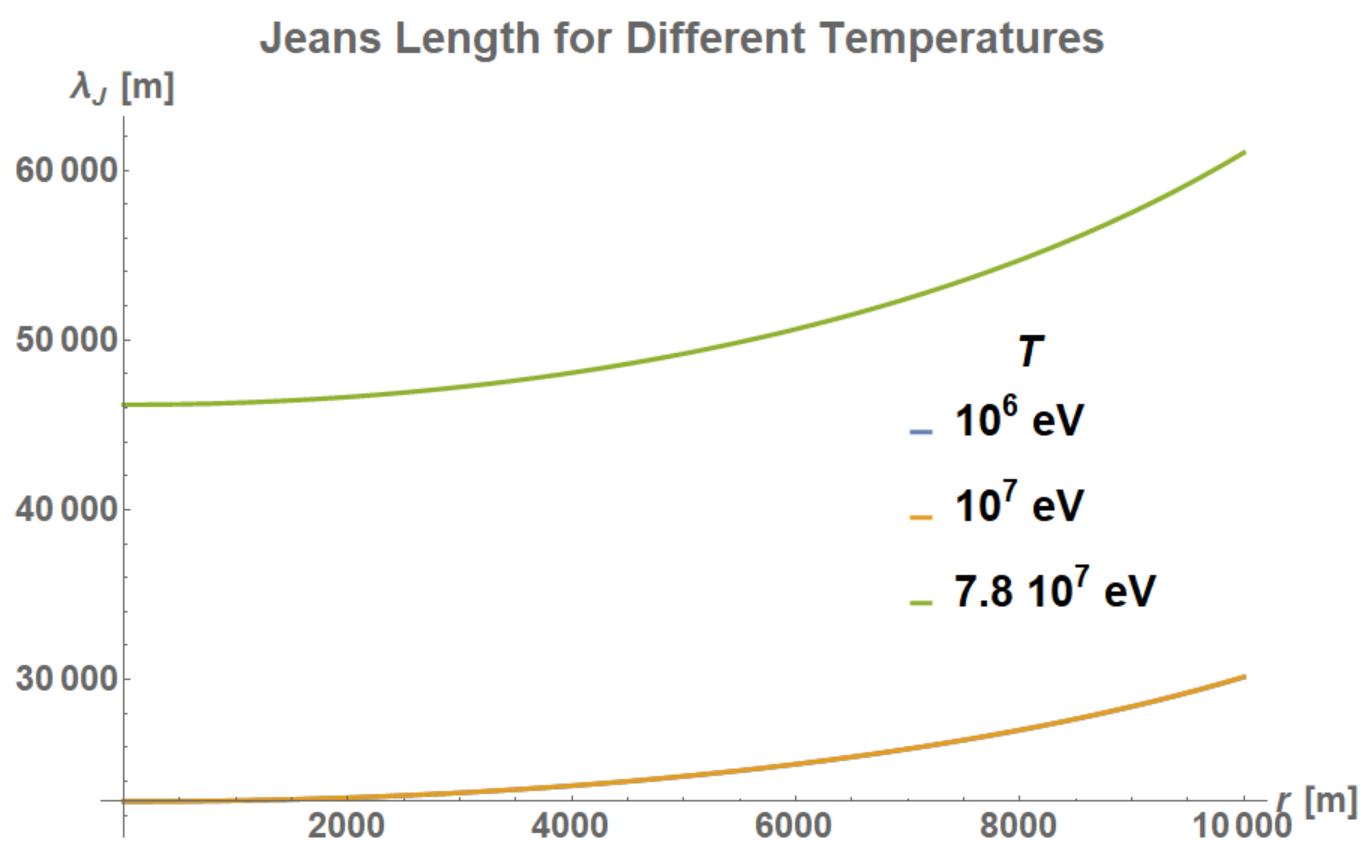}
    \hspace{5mm}
     \includegraphics[width=0.3\textwidth]{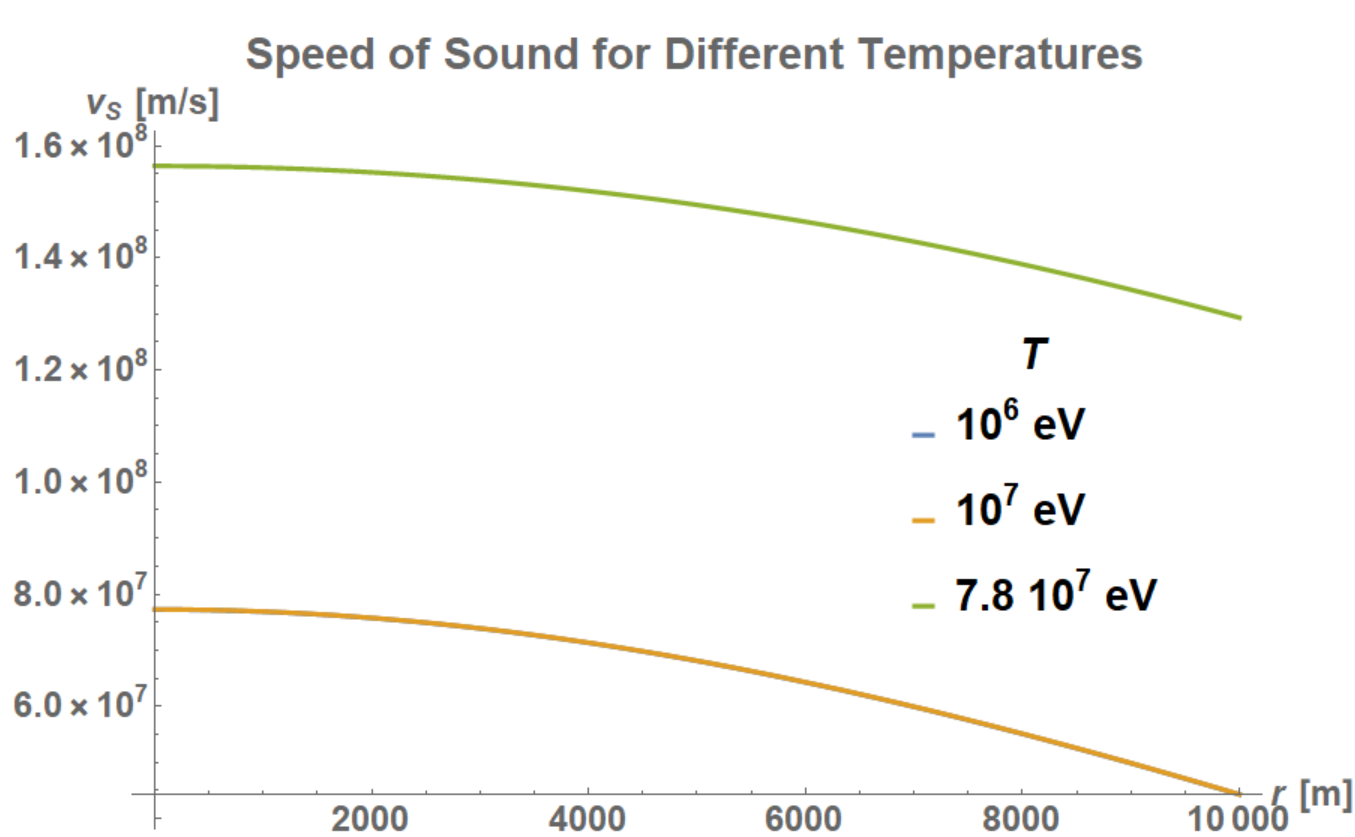}
        \caption{Like in Fig.~\ref{05rho0}, but keeping the central density equal to  $2\rho_0$.}  \label{2rho0}
 \end{figure}

 We solve numerically the TOV equations for $10^6\lesssim T<T_{\rm{max}}$ in the absence and in the presence of axions and plot the ratio between the mass of the star without axions $M_{\rm{s}}^*(r)$ over the mass of the star with axions $M_{\rm{ax}}^*(r)$, keeping the radius fixed -- $R=10$km, which is a typical size of neutron stars. In Figs.~\ref{05rho0},~\ref{rho0} and~\ref{2rho0} we present plots, where the central density is 
$\rho(0)=0.5\rho_0$, $\rho(0)=\rho_0$ and $\rho(0)=2\rho_0$, respectively.

 In so doing we confirm that, for the case of trapped axions, there is a temperature range for which the axions can influence the inner state of the star. For low-enough temperatures their effect is negligible, and for high-enough temperatures (proto neutron stars or merging neutron stars) they are  restricted by causality. For temperatures approaching the causal bound, but still definitely compatible with that bound,
 and for densities still much less than the critical density, the axions can have a conspicuous effect. 
 
From the presented plots we deduce that, for physical solutions of the TOV equations,
the ALPs increase the mass of the star.

Currently, there exists the so-called hyperon puzzle~\cite{hyppuzzle}, which consists in the fact that when adding hyperons to the matter content of the neutron star, the maximally attainable mass obtained from a modified equation of state, becomes smaller than what we observe in reality when the neutron star mass is around 2 solar masses -- e.g.~the mass of the pulsar PSR J0740+6620 can have such a mass~\cite{Salmi:2024aum}. We show that the presence of the hypothetical dark ALPs has the potential to address the existing discrepancy between the maximally observed neutron star masses and the maximal masses allowed by the theory of gravity, combined with all strong interaction nuclear effects. But once the interaction constant goes below some lower threshold value, the influence of the ALPs on the neutron-star mass becomes insignificant and unobservable. 


\section{Conclusion}
In this work we presented a model which accounts for the  pressure created by trapped axion-like particles (ALPs) in the interior of a neutron star. We did not consider strong interaction effects or exotic extensions (e.g.~quark gluon plasma~\cite{qgp}) other than the direct interaction between a light pseudo-scalar particle and neutrons. Axion-fermion interaction is independent of other forms of matter that might be present in a neutron star, therefore, we believe that our conclusions can be applied to specific models. 

The ALPs are produced through bremsstrahlung, and we derive the equation of state for the axionic gas that is present inside the star. This effect is added to the degeneracy pressure of neutrons and we computed the ratio between the mass of a pure neutron star over the mass of the star including this additional interaction. The trapping regime is restricted by causality and we show that there exists a region of allowed values of the free parameters, for which axions can have a significant effect on the mass of 
 neutron stars. For other central densities and temperatures   axions do not bring any noticeable effect to the structure of a neutron star. Our results 
 provide an interesting scenario to be tested through experiments. An advantage is the fact that the results can be realized in specific axion models, such as KSVZ or DFSZ~\cite{ksvz1, ksvz2, dszf1, dszf2} like models by choosing specific values of the parameters.

\section{Acknowledgements} 	
M.N. is partially supported by the Bulgarian Ministry of Education FNI-SU through the project 80-10-64/27.05.2025. M.N. would also like to thank to D. Mihaylov, S. Yazadjiev, T. Vetsov and V. Kozhuharov for fruitful discussions.

\end{document}